\documentclass[showpacs,preprintnumbers,superscriptaddress]{revtex4}
\usepackage{CJK}
\usepackage{amsmath,amssymb,graphicx,bm}
\begin{document}

\title{Entropy and Energy of Static Spherically Symmetric Black Hole in $f(R)$ theory}

\author{Yaoguang Zheng}
%\email{hesoyam12456@163.com}
\affiliation{College of Physical Science and Technology, Hebei University, Baoding 071002, China}
\author{Rong-Jia Yang \footnote{Corresponding author}}
\email{yangrongjia@tsinghua.org.cn}
\affiliation{College of Physical Science and Technology, Hebei University, Baoding 071002, China}
\affiliation{Hebei Key Lab of Optic-Electronic Information and Materials, Hebei University, Baoding 071002, China}

\begin{abstract}
We consider the new horizon first law in $f(R)$ theory with general spherically symmetric black hole. We derive the general formulas to computed the entropy and energy of the black hole. For applications, some nontrivial black hole solutions in some popular $f(R)$ theories are investigated, the entropies and the energies of black holes in these models are first calculated.
\end{abstract}


\maketitle

\section{Introduction}
There is a deep connection between thermodynamics and gravity. For a black hole, its area can be regarded as the entropy \cite{Bekenstein:1973ur},  and four laws for dynamics like in thermodynamics were suggested \cite{Bardeen:1973gs}. For diffeomorphism invariance of gravitational theory, the entropy of a black hole can be seen as a Noether charge \cite{Wald:1993nt, Iyer:1994ys}. The Einstein equations had been derived from the first law of thermodynamics \cite{Jacobson:1995ab}, which was generalized to the non-equilibrium thermodynamics of spacetime \cite{Eling:2006aw}. This connection was also investigated in modified gravity theories¡ªsuch as Lancos-Lovelock gravity \cite{Paranjape:2006ca}, $f(R)$ theory \cite{Elizalde:2008pv}, and the scalar-Gauss-Bonnet gravity \cite{Bamba:2009gq}. It was shown that the equation of motion for generalized gravity theory is equivalent to the thermodynamical relationship $\delta Q=T\delta S$ \cite{Brustein:2009hy}. In spherically symmetric spacetime Padmanabhan presented a general form for understanding the thermodynamics of the horizon \cite{Padmanabhan:2002sha}. For a time-dependent evolution horizon or a fixed axis-symmetric horizon, the Einstein equations on horizon can be rewritten as a thermodynamic identity \cite{Kothawala:2007em}. In cubic and quartic quasi-topological gravity, the field equations on $n+1$  dimensional topological black holes
horizon with constant curvature can be expressed like the form of the first thermodynamical law \cite{Sheykhi:2014rka}. In \cite{Yang:2011sx, Yang:2014kna}, one can obtain the thermal entropy density of any spacetime from Einstein equations without assuming the temperature or the horizon.

In higher-order theories of gravity, issues about the entropy and energy of black hole are important. Especially the problems related to the energy of black hole are problematic, some attempts have been made to find a satisfactory answer to this question \cite{Cognola:2011nj, Deser:2002jk, Deser:2007vs, Abreu:2010sc, Cai:2009qf}. Using the first law of the horizon, one can derived the entropy and energy of black hole in Einstein gravity \cite{Padmanabhan:2002sha}. Recently, as shown in \cite{Zheng:2018fyn}, the entropy and energy of black hole in $f(R)$ theories can be obtained simultaneously though the new horizon first law which was proposed in \cite{Hansen:2016gud}. Here we will consider the more complicated case: for the general spherically symmetric black hole we hope to obtain the energy and the entropy of black hole in $f(R)$ theory, which can reduce to the results in \cite{Zheng:2018fyn, Cognola:2011nj} for some special cases.

The rest of this article is organized as follows. In Sec. II, we briefly review the new horizon first law. In Sec. III, we investigate the entropy and energy of black holes in $f(R)$ Theories. In Sec. IV, we calculate the energies and entropies for black holes in some $f(R)$ theories for applications. Conclusion and discussion are given in Sec. V.

\section{The new horizon first law}
Usually if the radial component of the stress-energy tensor acts as the thermodynamic pressure, $P=T^{r}_{~r}|_{r_{+}}$, which is a special case of the assumption firstly proposed in \cite{Yang:2014kna}, and assume an horizon equation of state $P=P(V, T)$ with the temperature identified as the Hawking temperature and a geometric volume $V=V(r_{+})$ assigned to the horizon \cite{Parikh:2005qs}, then the radial Einstein equation can be written as the first law of thermodynamics, which can be rewritten as a horizon first law by considering a imaginary displacement of the horizon \cite{Padmanabhan:2002sha}
\begin{eqnarray}
\label{HFL}
\delta E=T\delta S-P\delta V,
\end{eqnarray}
where $E$ is the quasilocal energy and $S$ is the horizon entropy of the black hole. In Einstein gravity, $E$ proves to be the Misner-Sharp energy \cite{Misner:1964je}. The horizon first law (\ref{HFL}) is a special case of the `unified first law' \cite{Hayward:1997jp}.

There are two problems in this procedure that are needed to be further examined. The first problem is that the thermodynamic variables was vague in the original derivation, which require further determination. Secondly because both $S$ and $V$ are functions of only $r_+$, which makes the horizon first law (\ref{HFL}) to be of a function only depending on $r_+$. As a matter of fact, the equation (\ref{HFL}) can be rewritten as $\delta E=\left(TS^\prime+PV^\prime\right)\delta r_+$, where the primes stands for the derivative with respected to $r_+$. This makes the terms `heat' and `work' confused and results to a `vacuum interpretation' of the first law (\ref{HFL}) \cite{Hansen:2016gud}. To avoid the problems mentioned above, a new horizon first law was proposed in \cite{Hansen:2016gud}, by varying the horizon equation of state with the temperature $T$ and the pressure $P$ as independent thermodynamic quantities
\begin{eqnarray}
\label{nhel}
\delta G=-S\delta T+V\delta P,
\end{eqnarray}
which obviously depends on both $T$ and $P$, meaning that $T$ and $P$ can vary independently. The energy, and therefore the standard horizon first law (\ref{HFL}) can be obtained via a degenerate Legendre transformation $E=G+TS-PV$. The Gibbs free energy $G$ and the horizon entropy $S$ are derived concepts for specified volume. This new horizon first law has practical utility and provides further evidence for the connection between gravity and thermodynamics.

In 4-dimensional Einstein gravity, we briefly review this approach to explain how it works \cite{Hansen:2016gud}. Considering the geometry of a static spherically symmetric black hole, which is given by
\begin{eqnarray}
\label{metric0}
ds^2=-N(r)dt^2+\frac{dr^2}{N(r)}+r^2d\Omega^2,
\end{eqnarray}
The horizon is local at $r=r_+$ where $N(r_+)=0$ with $N'(r_+)\neq 0$.

Supposing minimal coupling to the matter and that the thermal sources are also the gravitational sources \cite{Yang:2014kna}, we have $P=T^r_{~r}$ and identify the Hawking temperature as the thermal temperature $T$ \cite{Padmanabhan:2002sha}, the $(^1_1)$ component of the Einstein equation can be written as at the horizon
\begin{eqnarray}
\label{P}
P=\frac{T}{2r_{+}}-\frac{1}{8\pi r^2_{+}},
\end{eqnarray}
which is the horizon equation of state, where we take the units $G=c=\hbar=1$.

Since the identification of the Hawking temperature as the thermal temperature does not fall back on any gravitational field equations \cite{Hansen:2016gud}. According to the conjecture suggested in \cite{Yang:2014kna}, the definition of the pressure is identified as the $(^r_{r})$ component of the matter stress-energy, which is also independent of any gravitational theories. So it is reasonable to speculate that the radial component of the gravitational field equations under consideration takes the following form \cite{Hansen:2016gud}
\begin{eqnarray}
\label{7}
P=D(r_+)+C(r_+)T,
\end{eqnarray}
where $C$ and $D$ are some functions of $r_+$, generally depending on the gravitational theory under consideration. Varying the generalized equation of state (\ref{7}) and multiplying the volume $V(r_+)$ of black hole, it is easy to get

\begin{eqnarray}
\label{9}
V\delta P=S\delta T+\delta G,
\end{eqnarray}
where
\begin{eqnarray}
\label{10}
G&=&\int^{r_+} V(r)D'(r)\,dr+T\int^{r_+} V(r)C'(r)dr\nonumber\\
&=&PV-ST-\int^{r_+} V'(r)D(r)dr,
\end{eqnarray}
and
\begin{eqnarray}
\label{11}
S=\int^{r_+} V'(r)C(r)dr,
\end{eqnarray}
by using the integration by parts. Taking the degenerate Legendre transformation $E=G+TS-PV$, we finally derive the energy as
\begin{eqnarray}
\label{energy}
E=-\int^{r_+} V'(r)D(r)dr.
\end{eqnarray}
Hypothesizing that $T$, $P$, and $V$ can be identified as the temperature, the pressure, and the volume, we can find that $G$, $S$, and $E$ are the Gibbs free energy, the entropy, and the energy of the black hole, respectively. For Einstein gravity in four dimensions, it is easily to have $C(r_+)=\frac{1}{2r_+}$ and $D(r_+)=-\frac{1}{8\pi r^2_+}$ from (\ref{P}), substituting them into Eq. (\ref{11}) and Eq. (\ref{energy}), we obtain $S=\pi r^2_+$ and $E=\frac{r_+}{2}$. It has been shown that the new horizon first law still works in $f(R)$ theories \cite{Zheng:2018fyn}.

\section{The entropy and energy of black hole in $f(R)$ theory}
We consider the general spherically symmetric and static black hole in $f(R)$ theories, whose geometry takes the form
\begin{eqnarray}
\label{metric}
ds^2=-W(r)dt^2+\frac{dr^2}{N(r)}+r^2 d\Omega^2,
\end{eqnarray}
where $W(r)$ and $N(r)$ are general functions of the coordinate $r$. Taking the largest positive root of $N(r_+)=0$, yields the event horizon which fulfils $N'(r_+)\neq 0$. The surface gravity is given by \cite{DiCriscienzo:2009hd}
\begin{eqnarray}
\kappa_K=\frac{\sqrt{W'(r_+) N'(r_+)}}{2},
\end{eqnarray}
which gives the temperature of the black hole as
\begin{eqnarray}
\label{temp}
T=\frac{\kappa_K}{2\pi}=\frac{\sqrt{W'(r_+)N'(r_+)}}{4\pi}.
\end{eqnarray}
In four-dimensional spacetime, the general action of $f(R)$ gravity theories with source is
\begin{eqnarray}
\label{12}
I=\int \text{d}^{4}x\sqrt{-g}\left[\frac{f(R)}{2k^2}+L_{\rm m}\right],
\end{eqnarray}
where $k^2=8\pi$, $f(R)$ is a function only depending on the Ricci scalar $R$, and $L_{\rm m}$ is the matter Lagrangian. Physically $f(R)$ must fulfil stability conditions \cite{Pogosian:2007sw}: (a) no ghosts, $df/dR>0$; (b) no tachyons, $d^2f/dR^2>0$ \cite{Dolgov:2003px}; (c) stable solutions, $df/dR/d^2f/dR^2>R$ \cite{Sawicki:2007tf}. Additionally $f(R)$ must also satisfy $\lim_{R\rightarrow \infty} (f(R)-R)/R=0$ for the existence of an effective cosmological constant at high curvature and $\lim_{R\rightarrow \infty} d(f(R)-R)/dR=0$ for recovering general relativity at early time. Varying the action (\ref{12}) with respect to the metric, yields the gravitational field equations
\begin{eqnarray}
\label{13}
G_{\mu\nu}\equiv R_{\mu\nu}-\frac{1}{2}g_{\mu\nu}R=k^2\left(\frac{1}{F}T_{\mu\nu}+\frac{1}{k^2}\mathcal{T}_{\mu\nu} \right),
\end{eqnarray}
where $F=\frac{df}{dR}$ and $T_{\mu\nu}=\frac{-2}{\sqrt{-g}}\frac{\delta L_{\rm m}}{\delta g^{\mu\nu}}$ is the energy-momentum tensor for the matter. The tress-energy tensor for the effective curvature fluid, $\mathcal{T}_{\mu\nu}$, is given as following
\begin{eqnarray}
\label{rtensor}
\mathcal{T}_{\mu\nu}=\frac{1}{F(R)}\left[\frac{1}{2}g_{\mu\nu}(f-RF)+\nabla_{\mu}\nabla_{\nu}F-g_{\mu\nu}\Box F \right],
\end{eqnarray}
where $\Box =\nabla ^{\gamma }\nabla _{\gamma}$. Taking the trace of the equation (13), gives the relation as follow
\begin{eqnarray}
\label{trace}
RF(R)-2f(R)+3\Box F(R)=k^2T.
\end{eqnarray}
For the general spherically symmetric and static black hole (\ref{metric}), the $(^1_1)$ components of the Einstein tensor and the stress-energy tensor of the effective curvature fluid are respectively given by
\begin{eqnarray}
\label{gtensor11}
G^{1}_{1}=\frac{1}{r^2}(-1+\frac{rNW'}{W}+N),
\end{eqnarray}
and
\begin{eqnarray}
\label{rtensor11}
\mathcal{T}^{1}_{1}=\frac{1}{F}\left[\frac{1}{2}(RF-f)+\frac{N}{2W}W'F'+\frac{2}{r}NF'\right],
\end{eqnarray}
where the prime denotes the derivative with respected to $r$. Substituting Eqs. (\ref{gtensor11}), (\ref{rtensor11}), and $T^1_1=P$ into the equation (\ref{13}), we derive
\begin{eqnarray}
\label{26}
k^2P=-\left[\frac{F}{r^2}+\frac{1}{2}(f-RF)\right]+\sqrt{\frac{N'}{W'}}\left(\frac{F}{r}+\frac{1}{2}F'\right)\sqrt{W'N'}+\frac{NF}{r^2}+\frac{2NF'}{r}.
\end{eqnarray}
At the horizon, $r=r_+$, thinking of the temperature (\ref{temp}) and $N(r_+)=0$, the equation (\ref{26}) reduces to
\begin{eqnarray}
\label{27}
P=-\frac{1}{8\pi}\left[\frac{F}{r^2_+}+\frac{1}{2}\left(f-RF\right)\right]+\frac{1}{4}\sqrt{\frac{N'}{W'}}\left(\frac{2F}{r_+}+F'\right)T.
\end{eqnarray}
Comparing Eq. (\ref{7}) with Eq. (\ref{27}), we then have
\begin{eqnarray}
\label{28}
D(r_+)=-\frac{1}{8\pi}\left[\frac{F}{r^2_+}+\frac{1}{2}\left(f-RF\right)\right],
\end{eqnarray}
and
\begin{eqnarray}
\label{29}
C(r_+)=\frac{1}{4}\sqrt{\frac{N'}{W'}}\left(\frac{2F}{r_+}+F'\right).
\end{eqnarray}
The volume $V$ of the black hole (\ref{metric}) is given by \cite{Parikh:2005qs}
\begin{eqnarray}
\label{V}
V(r_+)&=&\int_0^{r_+}\int_0^{\pi}\int_0^{2\pi}\sqrt{-g}\;dr d\theta d\phi \nonumber \\
&=&4\pi\int^{r_+}_{0}\sqrt{\frac{W'}{N'}}r^2dr.
\end{eqnarray}
where we have used the relation $\frac{N(r_+)}{W(r_+)}=\frac{N'(r_+)}{W'(r_+)}$ \cite{DiCriscienzo:2009hd}. Substituting Eqs. (\ref{V}) and (\ref{26}) into Eq. (\ref{11}), we obtain the entropy as
\begin{eqnarray}
\label{entropy1}
S&=&\int^{r_+} (2\pi rF+\pi r^2 F')dr=\pi r^{2}_{+}F,
\end{eqnarray}
thought the black hole (\ref{metric}) has different temperature and volume, its entropy (\ref{entropy1}) has the same form for the entropy of black hole with $W(r)=N(r)$ obtained in \cite{Zheng:2018fyn}. The formula (\ref{entropy1}) can be obtained by using the Euclidean semiclassical approach or the Wald entropy formula \cite{Dyer:2008hb, Vollick:2007fh, Iyer:1995kg}. Using the degenerate Legendre transformation $E=G+TS-PV$, inserting equations (\ref{V}) and (\ref{28}) into Eq. (\ref{energy}), then the energy of the black hole is given as follow
\begin{eqnarray}
\label{energy1}
E=\frac{1}{2}\int^{r_+} \sqrt{\frac{W'}{N'}}\left[\frac{F}{r^2}+\frac{1}{2}(f-RF)\right]r^2dr.
\end{eqnarray}
For $W(r)=N(r)e^{2\alpha(r)}$, the equation (\ref{energy1}) reduces to the results obtained in \cite{Cognola:2011nj} where the entropy was obtained by using the Wald method. For $W(r)=N(r)$, the equation (\ref{energy1}) gives the result obtained in \cite{Zheng:2018fyn}. Eqs. (\ref{entropy1}) and (\ref{energy1}) show that the new horizon first law still work in $f(R)$ theories for the general spherically symmetric and static black hole.
\section{Applications}
In this section, we will use Eqs. (\ref{entropy1}) and (\ref{energy1}) to calculate the entropies and energies of black holes (\ref{metric}) in some popular $f(R)$ theories to illustrate the method described above. These models allowe solutions with constant Ricci curvature (such as Schwarzschild-de Sitter or Schwarzschild solutions) or solutions with non-constant Ricci curvature. For a Schwarzschild-de Sitter solution, we have $N(r_+)=1-2M/r-R_{0}r^2/12=0$ at the horizon \cite{Multamaki:2006zb}, which gives
\begin{eqnarray}
\label{hc}
2M=r_{+}-\frac{1}{12}R_{0}r^3_{+}.
\end{eqnarray}
This equation relates the mass $M$, the radius $r_+$, and the Ricci curvature $R_0$ together.

We firstly start with a $f(R)$ model which unifies inflation and cosmic acceleration under the same picture and was confirmed by the solar system tests \cite{Nojiri:2007as}. This model takes the form
\begin{eqnarray}
f(R) = R- f_0 \int_0^R e^{-\alpha
\frac{R_1^{2n}}{\left(x-R_1\right)^{2n}}-f_0 \frac{x}{\Lambda_i}}
dx,
\end{eqnarray}
where $R_1$ is a constant which is given by $ f_0 R_1 \int_0^1
e^{-\alpha/x^2} dx  = R_{\rm now}$ with $0<f_0<1$ and $R_{\rm now}$ the Ricci scalar at present. The effective cosmological constant is $-f(-\infty) = \Lambda_i$ in the early universe and is $2R_0$ in the present era. The stability condition, $f''(R)>0$, gives
\begin{eqnarray}
\label{cond2}
\frac{f_0}{\Lambda_i} > \frac{2n R_1^{2n}}{\left(R - R_1
\right)^{2n+1}}.
\end{eqnarray}
Due to the fact that $f(0)=0$, this model allows a Schwarzschild solution. For a Schwarzschild-de Sitter solution, we obtain from Eq. (\ref{trace}),
\begin{eqnarray}
\label{40}
R_0+f_0 R_0 e^{-\alpha\frac{R_1^{2n}}{\left(R_0-R_1\right)^{2n}}-f_0 \frac{R_0}{\Lambda_i}}=2f_0 \int_0^R e^{-\alpha\frac{R_1^{2n}}{\left(x-R_1\right)^{2n}}-f_0 \frac{x}{\Lambda_i}}dx.
\end{eqnarray}
For the black hole (\ref{metric}), we calculate the entropy from Eq. (\ref{entropy1}) as
\begin{eqnarray}
S=\pi r_+^2\left[1- f_0e^{-\alpha
\frac{R_1^{2n}}{\left(R-R_1\right)^{2n}}-f_0 \frac{R}{\Lambda_i}}\right].
\end{eqnarray}
The nonnegativity of the entropy gives additional limits on the parameters: $1> f_0e^{-\alpha
\frac{R_1^{2n}}{\left(R-R_1\right)^{2n}}-f_0 \frac{R}{\Lambda_i}}$. The entropy for Schwarzschild-de Sitter black hole is
\begin{eqnarray}
\label{39}
S = \pi r_+^2\left[1- f_0e^{-\alpha
\frac{R_1^{2n}}{\left(R_0-R_1\right)^{2n}}-f_0 \frac{R_0}{\Lambda_i}}\right],
\end{eqnarray}
where $R_0$ must fulfil the equations (\ref{cond2}) and (\ref{40}).

For Schwarzschild black hole ($R_0=0$), the stability condition (\ref{cond2}) reduces to $n < \frac{f_0 R_1}{2\Lambda_i}$, and the entropy (\ref{39}) reduces to
\begin{eqnarray}
\label{ss2}
S=\pi r_+^2\left(1-f_0e^{-\alpha}\right).
\end{eqnarray}
Computing the energy by using Eq. (\ref{energy1}), we have
\begin{eqnarray}
\label{m2e}
E=\frac{1}{4}\int^{r_+}\sqrt{\frac{W'}{N'}}\left[2+f_0\left(R r^2-2\right)e^{-\alpha
\frac{R_1^{2n}}{\left(R-R_1\right)^{2n}}-f_0 \frac{R}{\Lambda_i}}-f_0r^2\int_0^{R}e^{-\alpha
\frac{R_1^{2n}}{\left(x-R_1\right)^{2n}}-f_0 \frac{x}{\Lambda_i}}dx\right]dr.
\end{eqnarray}
For the Schwarzschild-de Sitter black hole, equation (\ref{m2e}) reduces to
\begin{eqnarray}
\label{2energy}
E&=&\frac{r_+}{12}\left[6+f_0\left(R_0r_+^2-6\right)e^{-\alpha
\frac{R_1^{2n}}{\left(R_0-R_1\right)^{2n}}-f_0 \frac{R_0}{\Lambda_i}}-f_0r_+^2\int_0^{R_0}e^{-\alpha
\frac{R_1^{2n}}{\left(x-R_1\right)^{2n}}+f_0 \frac{x}{\Lambda_i}}dx\right]\\\nonumber
&=&\frac{r_+}{12}\left[6-\frac{1}{2}R_{0}r^2_{+}+\left(\frac{1}{2}R_0r_+^2-6\right)f_{0}e^{-\alpha
\frac{R_1^{2n}}{\left(R_0-R_1\right)^{2n}}-f_0 \frac{R_0}{\Lambda_i}}\right]\\\nonumber
&=&\left[1-f_{0}e^{-\alpha
\frac{R_1^{2n}}{\left(R_0-R_1\right)^{2n}}-f_0 \frac{R_0}{\Lambda_i}}\right]M,
\end{eqnarray}
where equations (\ref{hc}) and (\ref{40}) were used. For the Schwarzschild black hole, $R_0=0$, equation (\ref{2energy}) reduces to
\begin{eqnarray}
\label{s2energy}
E=\frac{r_+}{2}\left(1-f_0e^{-\alpha}\right).
\end{eqnarray}
Since $S\geq 0$ and $E\geq 0$, Eqs. (\ref{ss2}) and (\ref{s2energy}) give new constraints on the parameters: (a) $1>f_{0}e^{-\alpha
\frac{R_1^{2n}}{\left(R_0-R_1\right)^{2n}}-f_0 \frac{R_0}{\Lambda_i}}$ for Schwarzschild-de Sitter black hole; and (b) $1\geq f_0e^{-\alpha}$ for Schwarzschild black hole.

The second $f(R)$ model we consider here was recovered from the entropy of black holes \cite{Caravelli:2010be}
\begin{eqnarray}
f(R)=R-qR^{\beta+1}\frac{\alpha\beta+\alpha+\beta\epsilon}{\beta+1}+q\epsilon R^{\beta+1}\ln\left(\frac{a_0^{\beta}R^{\beta}}{c}\right),
\end{eqnarray}
where $0\leq\epsilon\leq \frac{e}{4}(1+\frac{4}{e}\alpha)$ and $q=4a_0^{\beta}/c(\beta+1)$ with $\alpha\geq 0$, $\beta>0$, $a_0=l_p^2$ and $c$ a constant. Since $R\neq 0$, this type of $f(R)$ theory admits no Schwarzschild solution. Use the equation (\ref{trace}), we get the condition for the Schwarzschild-de Sitter solution
\begin{eqnarray}
\label{42}
1+\beta-q\alpha R_0^{\beta}+q\alpha\beta R_0^{\beta}-2q\beta\epsilon R_0^{\beta}=\left(\beta^2-1\right)q\epsilon R_0^{\beta}\ln\left(\frac{a_0^{\beta}R_0^{\beta}}{c}\right).
\end{eqnarray}
From the Eq. (\ref{entropy1}), the entropy of the black hole (\ref{metric}) is computed as
\begin{eqnarray}
S=\pi r_+^2\left[1-\left(\beta+1\right)q \alpha R^{\beta}+\left(\beta+1\right)q \epsilon R^{\beta}\ln\left(\frac{a_0^{\beta}R^{\beta}}{c}\right)\right],
\end{eqnarray}
The nonnegativity of the entropy gives additional constraints on the parameters: $1>\left(\beta+1\right)q \alpha R^{\beta}+\left(\beta+1\right)q \epsilon R^{\beta}\ln\left(\frac{a_0^{\beta}R^{\beta}}{c}\right)$. For Schwarzschild-de Sitter black hole, the entropy is
\begin{eqnarray}
S&=&\pi r_+^2\left[1-\left(\beta+1\right)q \alpha R_0^{\beta}+\left(\beta+1\right)q \epsilon R_0^{\beta}\ln\left(\frac{a_0^{\beta}R_0^{\beta}}{c}\right)\right]\nonumber\\
&=&\frac{2\beta-q\beta(\alpha\beta-\alpha+2\epsilon)R_0^{\beta}}{\beta-1}\pi r_+^2\nonumber\\
&=&\frac{\beta\left[2-(\alpha\beta-\alpha+2\epsilon)q R_0^{\beta}\right]}{\beta-1}\pi r_+^2,
\end{eqnarray}
where the condition (\ref{42}) was uesed. Since $S\geq 0$, gives new limits on the parameters: $2>(\alpha\beta-\alpha+2\epsilon)q R_0^{\beta}$ for $\beta>1$ and $2<(\alpha\beta-\alpha+2\epsilon)q R_0^{\beta}$ for $\beta<1$.

The energy of the black hole (\ref{metric}) is derived from Eq. (\ref{energy1}) as
\begin{eqnarray}
\label{m3e}
E&=&\frac{1}{4}\int^{r_+}\sqrt{\frac{W'}{N'}}\left\{r^2 R\left[1-\frac{\alpha+\alpha\beta+\beta\epsilon}{\beta+1}+q\epsilon R^{\beta}\ln\left(\frac{a_0^{\beta}R^{\beta}}{c}\right)\right]
\right. \nonumber\\
&&\left.-\left(r^2 R-2\right)\left[1-\left(\beta+1\right)q\alpha R^{\beta}+\left(\beta+1\right)q\epsilon R^{\beta}\ln\left(\frac{a_0^{\beta}R^{\beta}}{c}\right)\right]\right\}dr\\\nonumber
&=&\frac{1}{4}\int\sqrt{\frac{W'}{N'}}\left\{2-2\alpha(\beta+1)q R^{\beta}-\frac{\alpha+\alpha\beta+\beta\epsilon}{\beta+1}r^2 R \right.\nonumber\\
&&\left.+q\alpha(\beta+1)r^2R^{\beta+1}+q\epsilon\left[2(\beta+1)-\beta r^2R\right]R^{\beta}\ln\left(\frac{a_0^{\beta}R^{\beta}}{c}\right)\right\}dr.
\end{eqnarray}
For Schwarzschild-de Sitter black hole, equation (\ref{m3e}) takes the form
\begin{eqnarray}
E&=&\frac{r_+}{12}\left\{q\beta R_0^{\beta}\left(r_+^2 R_0-12\right)\left[\alpha+\alpha\beta-\epsilon-\epsilon(\beta+1)\ln\left(\frac{a_0^{\beta}R_0^{\beta}}{c}\right)\right]\right\}\\\nonumber
&=&-2q\beta R_0^{\beta}M\left[\alpha+\alpha\beta-\epsilon-\epsilon(\beta+1)\ln\left(\frac{a_0^{\beta}R_0^{\beta}}{c}\right)\right]\\\nonumber
&=&\frac{2\beta\left[1+\beta-(\beta\epsilon+\alpha\beta^2+\epsilon-\alpha\beta)qR_0^{\beta}\right]}{\beta-1}M
\end{eqnarray}
where the equations (\ref{42}) and (\ref{hc}) were used. The nonnegativity of the energy gives new constraints on the parameters: $1+\beta>(\beta\epsilon+\alpha\beta^2+\epsilon-\alpha\beta)qR_0^{\beta}$ for $\beta>1$ and $1+\beta<(\beta\epsilon+\alpha\beta^2+\epsilon-\alpha\beta)qR_0^{\beta}$ for $\beta<1$.

We thirdly investigate the following $f(R)$ theory which allows black hole solutions with non-constant Ricci curvature \cite{Canate:2015dda}
\begin{eqnarray}
\label{49}
f(R)=2a\sqrt{R-\alpha},
\end{eqnarray}
where $\alpha$ is a parameter of the model which is related to an effective cosmological constant and $a > 0$ is a parameter with units [distance]$^{-1}$. This model admits a static spherically symmetric solution (\ref{metric}) which takes the follow form \cite{Canate:2015dda}
\begin{eqnarray}
\label{50}
N(r)=W(r)=\frac{1}{2}\left(1-\frac{\alpha r^2}{6}+\frac{2Q}{r^2}\right),
\end{eqnarray}
where $Q$ is an integration constant. The event horizon of the black hole is located at: (a) $r_+=\sqrt{3\alpha+\alpha\sqrt{9+12\alpha Q}}/\alpha$ for $\alpha>0$ and $Q>0$; (b) $r_+=\sqrt{3\alpha-\alpha\sqrt{9+12\alpha Q}}/\alpha$ for $\alpha>0$, $Q<0$ and $\alpha Q>-3/4$; (c) $r_+=\sqrt{3/\alpha-\sqrt{9+12\alpha Q}/\alpha}$ for $\alpha<0$ and $Q<0$; and (d) $r_+=\sqrt{6/\alpha}$ for $\alpha>0$ and $Q=0$. The Ricci scalar evolves as
\begin{eqnarray}
\label{51}
R=\alpha+\frac{1}{r^2},
\end{eqnarray}
Substituting $W'(r_+)=N'(r_+)$ into Eq. (\ref{entropy1}), the entropy is calculated as
\begin{eqnarray}
\label{s41}
S=\frac{a\pi(3+\sqrt{9+12\alpha Q})}{\alpha}, ~~~~{\rm for}~~r_+=\frac{\sqrt{3\alpha+\alpha\sqrt{9+12\alpha Q}}}{\alpha},
\end{eqnarray}
%%%%%%%%%%%%%%
\begin{eqnarray}
\label{s42}
S=\frac{a\pi(3-\sqrt{9+12\alpha Q})}{\alpha}, ~~~~{\rm for}~~r_+=\frac{\sqrt{3\alpha-\alpha\sqrt{9+12\alpha Q}}}{\alpha},
\end{eqnarray}
%%%%%%%%%%%%%%%%%%%%
\begin{eqnarray}
\label{s43}
S=\frac{a\pi(3-\sqrt{9+12\alpha Q})}{\alpha}, ~~~~{\rm for}~~r_+=\sqrt{\frac{3}{\alpha}-\frac{\sqrt{9+12\alpha Q}}{\alpha}},
\end{eqnarray}
%%%%%%%%%%%%%%%%%%%%%%
\begin{eqnarray}
\label{s44}
S=\frac{6a\pi}{\alpha}, ~~~~{\rm for}~~r_+=\sqrt{\frac{6}{\alpha}}.
\end{eqnarray}
From Eq. (\ref{energy1}), the energy is given by
\begin{equation}
E=-\frac{a}{12\alpha}(3+\sqrt{9+12\alpha Q})(\sqrt{9+12\alpha Q}-4), ~~~~{\rm for}~~r_+=\frac{\sqrt{3\alpha+\alpha\sqrt{9+12\alpha Q}}}{\alpha},
\end{equation}
%%%%%%%%%%%%%%%%%%
\begin{equation}
E = \frac{a}{12\alpha}(3-\sqrt{9+12\alpha Q})(4+\sqrt{9+12\alpha Q}), ~~~~{\rm for}~~r_+=\frac{\sqrt{3\alpha-\alpha\sqrt{9+12\alpha Q}}}{\alpha},
\end{equation}
\begin{equation}
E =\frac{a}{12\alpha}(3-\sqrt{9+12\alpha Q})(4+\sqrt{9+12\alpha Q}), ~~~~{\rm for}~~r_+=\sqrt{\frac{3}{\alpha}-\frac{\sqrt{9+12\alpha Q}}{\alpha}},
\end{equation}
\begin{equation}
E=\frac{a}{2\alpha}, ~~~~{\rm for}~~r_+=\sqrt{\frac{6}{\alpha}}.
\end{equation}
For $\alpha>0$ and $Q>0$, the nonnegativity of the energy gives additional constraints on the parameters: $\alpha Q<7/12$.

We finally consider a power-law of $F(r)$, for example: $F=\alpha r^a$, with constants $a$ and $\alpha$. The functions $W(r)$ and $N(r)$ in (\ref{metric}) are found to take forms \cite{Amirabi:2015aya}
\begin{eqnarray}
\label{CW}
W=r^{\frac{2a(a-1)}{a+2}}N
\end{eqnarray}
and
\begin{eqnarray}
\label{CN}
N=C_1r^{-\frac{2a^2+2a+2}{a+2}}+\frac{(a+2)^2}{(2a^2+2a+2)(2+2a-a^2)}.
\end{eqnarray}
where $C_1$ is a integration constant. The event horizon is local at $r_+=\left[-\frac{(2+a)^2}{C_1(2a^2+2a+2)(2+2a-a^2)} \right] ^{-\frac{2+a}{2a^2+2a+2}}$. In this case, function $f(R)$ is given by
\begin{eqnarray}
f(R)=\alpha_1R^{1-\frac{a}{2}},
\end{eqnarray}
where $\alpha_1=2\alpha(2-a)^{\frac{a}{2}-1}\left(\frac{3a}{2+2a+a^2}\right)^{\frac{a}{2}}$. The entropy (\ref{entropy1}) and the energy (\ref{energy1}) respectively reads
\begin{eqnarray}
\label{S44}
S&=&\alpha\pi r_{+}^{a+2}\nonumber\\
&=&\alpha\pi\left[-\frac{C_1(2a^2+2a+2)(2+2a-a^2)}{(2+a)^2} \right]^{\frac{(2+a)^2}{2a^2+2a+2}},
\end{eqnarray}
%%%%%%%%%%%%%%%%%%%
\begin{eqnarray}
\label{E44}
E&=&\frac{\pi^\frac{1}{2}}{8\Gamma(\frac{3}{2})}\dfrac{\alpha(a+2)^3 r_+^{\frac{2a^2+2a+2}{a+2}}}{(a^2-2a-2)(2a^2+2a+2)}\nonumber\\
&=&\frac{1}{4}\alpha C_1(2+a),
\end{eqnarray}
For $a=0$, we obtain the results in Einstein's gravity. The nonnegativity of the entropy and the energy give new constraints on the parameters: $\alpha\geq 0$ and $C_1(a-2)\geq 0$.

\section{Conclusions and discussions}
We investigated whether the new horizon first law still holds in $f(R)$ theory with general spherically symmetric black hole (\ref{metric}). We derived the general formulas to computed the entropy and energy of the black hole. For black hole (\ref{metric}), its temperature, volume and entropy have different forms, but its entropy (\ref{entropy1}) has the same form for the entropy of black hole (\ref{metric0}) obtained in \cite{Zheng:2018fyn}. For applications, some nontrivial black hole solutions in some popular $f(R)$ theories are considered, the entropies and the energies of black holes in these models are first calculated. The nonnegativity of entropy or energy gives new constraints on the parameters of $f(R)$ theories, which may be useful for future researches. The formulas presented here can be used to calculate the energies and entropies of black holes in other types $f(R)$ theories, but cannot be used in the case discussed in \cite{Nashed:2019tuk}, where $F(R)=0$.  In Reference \cite{Bamba:2009id}, a picture of equilibrium thermodynamics on the apparent horizon in the expanding cosmological background was obtained for a wide class of modified gravity theories. Whether this procedure can be applied to the apparent horizon in the expanding cosmological background worths further study

\begin{acknowledgments}
This study is supported in part by National Natural Science Foundation of China (Grant No. 11273010), the Hebei Provincial Outstanding Youth Fund (Grant No. A2014201068), the Outstanding Youth Fund of Hebei University (No. 2012JQ02).
\end{acknowledgments}

\bibliographystyle{ieeetr}%{elsarticle-num}
\bibliography{ref}

\end{document}